\documentclass[aps,prl,reprint,superscriptaddress]{revtex4-2}

\newcommand{\editor}[2]{%
  \expandafter\newcommand\csname #1note\endcsname[1]{%
    \textcolor{#2}{(\textbf{#1:} ##1)}}%
  \expandafter\newcommand\csname #1\endcsname[1]{%
    \textcolor{#2}{##1}}%
  \expandafter\newcommand\csname #1cancel\endcsname[1]{%
    \textcolor{#2}{\sout{##1}}}%
  \expandafter\newcommand\csname #1change\endcsname[2]{%
    \textcolor{#2}{\sout{##1} ##2}}%
  \newenvironment{#1text}{\color{#2}}{\color{black}}
}

\newcommand{\hh}{\mathcal{H}}
\newcommand{\ha}{\mathcal{A}}

\editor{ED}{purple}

\usepackage[T1]{fontenc}
\usepackage[utf8]{inputenc} 
\usepackage[english]{babel} 
\usepackage{orcidlink}
\usepackage{amsmath}
\usepackage{physics}
\usepackage{lipsum}
\usepackage{bm}

\begin{document}

\title{Spin Dynamics from Atomistic Quantum Simulations}


\author{Enrico Drigo\,\orcidlink{0000-0002-1797-2987}}
\email[]{endrigo@uchicago.edu}
\affiliation{Pritzker School of Molecular Engineering, University of Chicago, Chicago, Illinois 60637, USA}
\author{Marquis M. McMillan\,\orcidlink{0009-0009-4664-6556}}
\affiliation{Pritzker School of Molecular Engineering, University of Chicago, Chicago, Illinois 60637, USA}
\author{Benjamin Pingault\,\orcidlink{0000-0003-1259-2747}}
\affiliation{Q-NEXT, Argonne National Laboratory, Lemont, Illinois 60439, USA}
\affiliation{Materials Science Division, Argonne National Laboratory, Lemont, Illinois 60439, USA}
\affiliation{Pritzker School of Molecular Engineering, University of Chicago, Chicago, Illinois 60637, USA}
\author{Yinan Dong\,\orcidlink{0000-0002-5996-642X}}
\affiliation{Pritzker School of Molecular Engineering, University of Chicago, Chicago, Illinois 60637, USA}
\author{F. Joseph Heremans\,\orcidlink{0000-0003-3337-7958}}
\affiliation{Pritzker School of Molecular Engineering, University of Chicago, Chicago, Illinois 60637, USA}
\affiliation{Q-NEXT, Argonne National Laboratory, Lemont, Illinois 60439, USA}
\affiliation{Materials Science Division, Argonne National Laboratory, Lemont, Illinois 60439, USA}
\author{David D. Awschalom\,\orcidlink{0000-0002-8591-2687}}
\email{awsch@uchicago.edu}
\affiliation{Pritzker School of Molecular Engineering, University of Chicago, Chicago, Illinois 60637, USA}
\affiliation{Q-NEXT, Argonne National Laboratory, Lemont, Illinois 60439, USA}
\affiliation{Materials Science Division, Argonne National Laboratory, Lemont, Illinois 60439, USA}
\affiliation{Department of Physics, University of Chicago, Chicago, Illinois 60637, USA}
\author{Giulia Galli}
\email[]{gagalli@uchicago.edu}
\affiliation{Pritzker School of Molecular Engineering, University of Chicago, Chicago, Illinois 60637, USA}
\affiliation{Department of Chemistry, University of Chicago, Chicago, Illinois 60637, USA}
\affiliation{Materials Science Division, Argonne National Laboratory, Lemont, Illinois 60439, USA}


\date{\today}

\begin{abstract}
Optically active solid-state spin defects are promising candidates for quantum applications, however a unified theoretical framework to predict their spin dynamics at high temperatures is not yet available.
Here, using Kubo linear--response theory, we derive expressions of spin-lattice and decoherence times \(T_1\) and \(T_2\) in terms of  correlation functions of spin--lattice couplings.
We then evaluate \(T_1\) and \(T_2\) from molecular dynamics  and spin--lattice interaction time--series generated by state--of--the--art machine learning models trained on {\it ab--initio} data. Finally we measure \(T_1\) times for the NV center in diamond and compare experimental and theoretical results, showing excellent agreement.

\end{abstract}


\maketitle

{\it Introduction.}---Spin defects in solids are promising platforms for advancing quantum technologies, 
and extending their quantum coherence to high temperatures is of great interest for quantum sensing and quantum communication applications.

Recently, record coherence times at ambient conditions have been reported for a prototypical spin defect, the NV center in diamond, using isotopically purified samples~\cite{abobeih_one-second_2018,bar-gill_solid-state_2013,liu_coherent_2019,herbschleb_ultra-long_2019}.
While it was shown that the NV spin can be optically polarized up to $\mathrm{\approx700~K}$, little is known about the physical processes that limit coherence at higher temperatures, and a unified theoretical framework to describe high--temperature spin relaxation is not yet available.

Several methods have been proposed to compute spin relaxation times, including cluster correlation expansion (CCE) techniques~\cite{yang_quantum_2009}, the Redfield master equation~\cite{redfield_theory_1965} and Kubo linear response theory (LRT)~\cite{kubo_general_1954}.
CCE approaches are widely used to quantify the dissipative dynamics of a qubit in a bath of nuclear spins~\cite{ye_spin_2019,onizhuk_bath-limited_2023,onizhuk_understanding_2024,onizhuk_probing_2021}.
The Redfield master equation provides a perturbative formalism for the evolution of the reduced--density matrix~\cite{redfield_theory_1965}, and has been successfully applied to  spin--lattice relaxation in two--level systems at low temperatures under the approximation of one and two--phonon processes~\cite{cambria_temperature-dependent_2023,mariano_role_2025,lunghi_toward_2022,Lunghi2017,Lunghi2023,Xu2020,Xu2024}. 

Kubo LRT describes the regression of conserved quantities, such as energy and magnetization, toward equilibrium and can account for the time evolution of lattice vibrations~\cite{kubo_statistical-mechanical_1957,kubo_general_1954}. 
In 1946, Bloch proposed a set of phenomenological equations for the time evolution of the spin magnetization, $\mathbf{M}$~\cite{bloch_nuclear_1946}:
\begin{equation}
    \dot{\mathbf{M}}(t)=\gamma\mathbf{M}(t)\times\mathbf{H}(t) -\overline{\mathbf{R}}\mathbf{M}(t),\label{eq:bloch_phenomenological_equations}
\end{equation}
where $\mathbf{H}$ is the external magnetic field, $\gamma$ is the gyromagnetic ratio, $\overline{\mathbf{R}}=\mathrm{diag}\left\{\frac{1}{T_2},\frac{1}{T_2},\frac{1}{T_1}\right\}$, and $T_2$ and $T_1$ are the transverse (decoherence) and longitudinal (spin--lattice) relaxation times, respectively. 
At finite temperatures, the irreversible energy exchange between spins and lattice causes longitudinal relaxation~\cite{bloch_nuclear_1946,hahn_spin_1950,kubo_general_1954}.
In a similar fashion, the coupling between lattice vibrations and magnetic degrees of freedom affects the precession of $\mathbf{M}$, leading to decoherence~\cite{hahn_spin_1950,kubo_general_1954}.

In this Letter, within Kubo LRT, we identify the external perturbations responsible for spin relaxation and derive general expressions for spin--lattice and decoherence times $T_1$ and $T_2$ as a function of temperature.
We then compute spin relaxation times of the NV center in diamond, in the absence of magnetic noise, across a broad range of thermodynamic conditions using machine--learning (ML) molecular dynamics (MD) simulations and neural--network (NN) models of spin--lattice interactions. The effect of magnetic noise on $T_2$ is included at T=0 using results obtained with the CCE approach.
We also measure $T_1$ as a function of temperature for NV centers in diamond and compare our numerical and experimental results, finding excellent agreement.

{\it Theoretical framework.}---We consider a spin--lattice system characterized by its total energy (E) and magnetization ($\mathbf{M}$), and subject to an external perturbation applied adiabatically.
In the presence of an external field, E and $\mathbf{M}$ differ from their respective equilibrium  values~\cite{kubo_statistical-mechanical_1957,kubo_general_1954}, and upon removal of the driving field, after reaching an out-of-equilibrium steady state, E and  $\mathbf{M}$ attain their equilibrium values~\cite{kubo_general_1954,heitler_time_1936,hahn_spin_1950}.
In the out--of--equilibrium steady state an energy imbalance is present between spins and lattice~\cite{heitler_time_1936,kubo_general_1954}, and spin--lattice and decoherence relaxation processes characterize the regression of energy and transverse magnetization toward equilibrium, respectively~\cite{heitler_time_1936,kubo_general_1954,hahn_spin_1950}. 

Spin and lattice degrees of freedom are described by the Hamiltonian $\hh$,
\begin{align}
    \hh=\hh_\mathrm{S}^0 + \hh_\mathrm{L}+\hh_{\mathrm{SL}},
\end{align}
where $\hh_\mathrm{S}^0$ is the frozen--lattice spin Hamiltonian, $\hh_\mathrm{L}$ is the lattice contribution, $\comm{\hh_\mathrm{S}^0}{\hh_\mathrm{L}}=0$ and $\hh_{\mathrm{SL}}$ is the interaction between magnetic and vibrational degrees of freedom. 
Introduced in Refs.~\cite{pryce_modified_1950,mcweeny_origin_1965}, spin Hamiltonians, $\hh_\mathrm{S}$, characterize perturbatively the effect of nuclear and electronic spins on the degenerate Born--Oppenheimer ground state. In general, $\hh_\mathrm{S}$ is a parametric function of the atomic positions and therefore has an implicit dependence on the lattice degrees of freedom. 
Following  Refs.~\cite{lasoroski_vibrational_2014,khan_ab_2017}, we define the frozen--lattice spin Hamiltonian $\hh_\mathrm{S}^0\equiv\expval{\hh_\mathrm{S}}_L$ and spin--lattice coupling $\hh_\mathrm{SL}\equiv\hh_\mathrm{S}-\expval{\hh_\mathrm{S}}_L$. $\expval{\cdot}_L$ is the expectation value computed over the lattice degrees of freedom.

In the limit of high temperatures and at second order in $\hh_\mathrm{SL}$, the Kubo LRT formulation of spin relaxation yields~\cite{kubo_general_1954}:
\begin{align}
    &\frac{1}{T_1}=\sum_{\omega_\delta\neq0}\omega_\delta^2\int_0^\infty\dd{t}\,e^{i\omega_\delta t}\frac{\expval{\hh^{(\delta)}_\mathrm{SL}(t)\hh^{(-\delta)}_\mathrm{SL}}}{\expval{{\hh_\mathrm{S}}^2}},\label{eq:spin-lattice relaxation}\\
    &\frac{1}{T_2}=\sum_{\omega_\delta}\int_0^\infty\dd{t}\,e^{i\omega_\delta t}\frac{\expval{\comm{\bm{\mathcal{M}}^\perp}{\hh^{(\delta)}_\mathrm{SL}(t)}\cdot\comm{\bm{\mathcal{M}}^\perp}{\hh^{(-\delta)}_\mathrm{SL}}}}{\hbar^{2}\expval{\abs{\bm{\mathcal{M}}^\perp}^2}}.\label{eq:decoherence relaxation}
\end{align}
where we note that according to Luttinger's theory of thermal transport, the perturbation inducing the out--of--equilibrium energy distribution is proportional to the spin Hamiltonian~\cite{luttinger_theory_1964,kubo_general_1954}.
In Eqs.\eqref{eq:spin-lattice relaxation}--\eqref{eq:decoherence relaxation}, $\bm{\mathcal{M}}^\perp$ is the transverse magnetization operator, $\expval{\cdot}$ is the expectation value computed over $\hh_0=\hh_\mathrm{S}^0+\hh_\mathrm{L}$, $\hh^{(\delta)}_\mathrm{SL}$ is the projection of $\hh_\mathrm{SL}$ between two eigenstates of $\hh_\mathrm{S}^0$, $(n,m)\equiv\delta$, and $\omega_\delta=\frac{E_n-E_m}{\hbar}$. 

In the interaction picture relative to $\hh_0$, the evolution of $\hh_\mathrm{SL}$ is separable between spins and lattice: $\hh_\mathrm{SL}(t)=\sum_\delta e^{i\omega_\delta t}\hh^{(\delta)}_\mathrm{SL}(t)$ where $\hh_\mathrm{SL}^{(\delta)}(t)\equiv e^{i\hh_\mathrm{L} t/\hbar}\hh_\mathrm{SL}^{(\delta)}e^{-i\hh_\mathrm{L} t/\hbar}$.
The time evolution of the lattice can be sampled via lattice dynamics or MD approaches depending on the temperature regime and level of accuracy of interest.
At low temperatures, phonon quasi particles accurately describe the crystal. At high temperatures, the dynamics of lattice vibrations can be estimated by integrating the classical equations of motion of the nuclei. In the following, we present a molecular dynamics approach based on ML force fields and NN models of the zero--field--splitting (ZFS) tensor to compute spin relaxation times at finite temperature, in the absence of a magnetic bath, with {\it ab--initio} accuracy.
We then carry out  measurements on the isotopically purified NV center in diamond and show that our theoretical and experimental results are in  excellent agreement.

{\it Results.}---For the NV center in diamond, the major contribution to the spin Hamiltonian comes from the ZFS  and can be written as: $\hh_\mathrm{ZFS}=\bm{\mathcal{S} }\overline{\bm{\mathcal{D}}}\bm{\mathcal{S} }^\mathrm{T}$, where $\bm{\mathcal{S} }$ is the spin vector operator and  $\overline{\bm{\mathcal{D}}}$ is the ZFS tensor~\cite{pryce_modified_1950,mcweeny_origin_1965,neese_calculation_1998,neese_importance_2006,neese_calculation_2007,rayson_first_2008}. In this system, spin--orbit coupling contributions to ZFS are negligible compared to spin-spin ones~\cite{neese_calculation_1998,neese_importance_2006,neese_calculation_2007,rayson_first_2008}.
The dipolar interaction leads to the splitting of the NV's triplet ground state, $^3A_2$, into the $\left\{\ket{0}, \ket{+}, \ket{-}\right\}$  sub--levels.
The first--principle (FP) evaluation of the ZFS tensors and of MD simulation trajectories has a high computational cost, as reliable estimates of spin relaxation times require long trajectories (on the order of ns) and large supercells with hundreds of atoms.
Hence, straightforward FP approaches are impracticable,  and here we rely on ML methods to accelerate both MD simulations and ZFS calculations, achieving {\it ab--initio} accuracy at a fraction of the computational cost.

To evaluate Eqs.~\eqref{eq:spin-lattice relaxation}--\eqref{eq:decoherence relaxation}, we train a MACE ML interatomic potential (MLIP) to approximate the FP potential energy surface, and a NN model to compute the ZFS tensors~\cite{NEURIPS2022_4a36c3c5}. 
The MACE MLIP is trained to reproduce {\it ab--initio} total energies and atomic forces computed  at the spin--polarized DFT/PBE level of theory~\cite{doi:10.1021/acs.jctc.3c00249,PhysRevLett.77.3865}. 
First, we run $\mathrm{2}$ to $\mathrm{10~ps}$ Born--Oppenheimer (BO) MD simulations of one NV center in  diamond cells of $64$ and $512$ atoms in the NVE and NVT ensembles in the $\mathrm{1000}$--$\mathrm{2500~K}$ temperature range~\cite{doi:10.1021/acs.jctc.3c00249}. 
Further, we apply an active learning scheme to optimally add structures to the training dataset using an uncertainty estimation approach~\cite{bidoggia2026aiidatrainspotautomatedtrainingneuralnetwork,SM}. 



All spin Hamiltonian calculations are performed using the pyZFS code and spin polarized wave--functions at the PBE level of theory~\cite{doi:10.1021/acs.jctc.3c00249,Ma2020,PhysRevLett.77.3865}.  
The ZFS NN model is a message passing equivariant graph NN based on the MACE architecture~\cite{NEURIPS2022_4a36c3c5,SM}. Equivariance is imposed exactly via contractions with Clebsh--Gordan coefficients matrices, ensuring that the final output transforms under rotations as a spherical harmonics of order 2~\cite{NEURIPS2022_4a36c3c5}.
Since the ZFS tensor depends only on the local environment of the defect, we enforce a local attention criterion using the Frobenius norm of the predicted atomic tensors $\overline{\bm{\mathcal{T}}}(\bm{R})$. 
We define the {\it center} as the atom with the highest attention.
The atoms within a given cutoff from the center are labeled {\it active}. The output of the NN is the sum of the contributions of the atoms in the active region, $\mathcal{A}$. 
The remaining atomic tensors are introduced in the loss function, $\mathcal{L}$, as a penalty. The loss function, over the training set, $\mathcal{D}$, is: $\mathcal{L}=\sum_{i\in\mathcal{D}}\norm{\overline{\mathbf{D}}_i-\sum_{j\in\mathcal{A}}\overline{\bm{\mathcal{T}}}(\bm{R}_j^i)}^2+\norm{\sum_{j\notin\mathcal{A}}\overline{\bm{\mathcal{T}}}(\bm{R}_j^i)}^2$.
Thanks to the attention criterion, the ZFS NN prediction is local. Indeed, the ZFS tensor depends only on the concentration of defects and its elements converge to finite values in the dilute limit.
The extension of the ZFS NN model to multiple spin defects is currently under investigation.

\begin{figure}[t]
    \centering
    \includegraphics[width=1\linewidth]{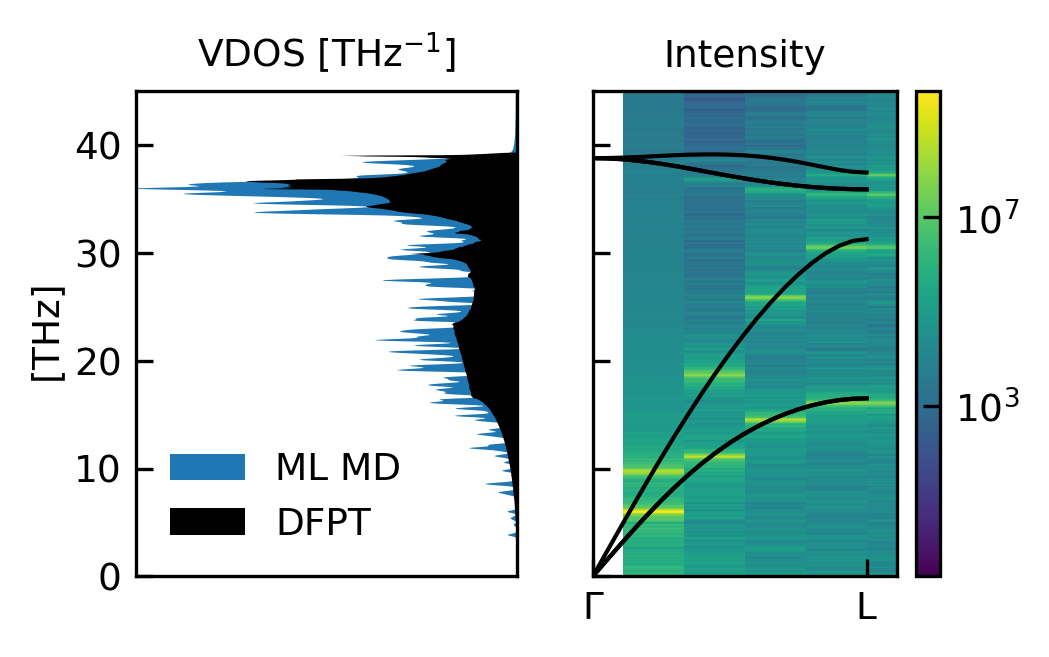}
    \caption{Left panel: comparison of the vibrational density of states (VDOS) of the NV center in diamond computed from molecular dynamics (MD) simulations with machine learned (ML) potentials in the dilute limit, and of crystalline diamond computed via density functional perturbation theory (DFPT)~\cite{baroni_phonons_2001}. Right panel: comparison of the phonon dispersion of the NV center in diamond computed from MD in the dilute limit and of crystalline diamond computed via DFPT~\cite{baroni_phonons_2001}. 
    The ML MD calculations are performed using a $\mathrm{2~ns}$ NVE trajectory of one NV center in a diamond cell of $5832$ atoms at $1000~\mathrm{K}$.
    }
    \label{fig: comparison ML DFPT}
\end{figure}
The MACE MLIP is benchmarked in the dilute limit against the vibrational density of states (VDOS) and phonon dispersion of bulk diamond (lattice constant =  $3.567~\text{\AA}$), obtained from density functional perturbation theory (DFPT)~\cite{baroni_phonons_2001}. 
We compute the ML VDOS and phonon dispersion from a $2~\mathrm{ns}$ ML MD trajectory at $\mathrm{1000~K}$ of one NV center in a diamond cell of $5832$ atoms harvesting configurations every timestep. 
In the quasi--harmonic approximation, the phonon dispersion is estimated from the poles of the Fourier transform of the time correlation function of the projection of the atomic displacements over a plane wave with wave--vector in the $\Gamma$--$\mathrm{L}$ direction of the crystal. 
The comparison between DFPT and ML MD results in the dilute limit, presented in Fig.~\ref{fig: comparison ML DFPT}, shows excellent agreement.

We also investigated the temperature dependence of the ZFS axial coefficient, $D$, from $500~\mathrm{ps}$--long NVE trajectories of one NV center in a diamond cell of $512$ atoms, and compared it with experimental results~\cite{acostaprl2010,liu_coherent_2019}.  
The $D$ coefficients are computed  using the ZFS NN model on configurations sampled every $\mathrm{10~fs}$. 
Our calculations show a relative shift of $\approx\mathrm{150~MHz}$ between the values of $D$ obtained from ML MD calculations and experiments. This difference does not reflect the inaccuracy of the ZFS NN model but rather  the limitations of the PBE functional used in computing the magnitudes of the elements of the ZFS tensor.
Importantly, the evaluation of $\frac{\partial D}{\partial T}$ at room temperature from ML MD and ZFS NN, $\approx\mathrm{-100~\frac{kHz}{K}}$, is in  good agreement with experimental findings, $\mathrm{-76(1)~\frac{kHz}{K}}$~\cite{acostaprl2010} and $\mathrm{-80~\frac{kHz}{K}}$~\cite{toyli_measurement_2012}.

After these successful benchmarks, we used the trained MACE MLIP and ZFS NN to carry out ns--long ML MD simulations and to evaluate the corresponding ZFS time--series of one NV center in diamond in a wide temperature interval spanning $400~$ to $1000~\mathrm{K}$.
Initially, the system is equilibrated in the NVT ensemble for $100~\mathrm{ps}$ and then $16~\mathrm{ns}$--long trajectories are collected in the NVE ensemble. 
From the ZFS time-series obtained with the MACE MLIP and the ZFS NN model, we compute spin--lattice and decoherence relaxation times at different temperatures. 
\begin{figure}[t]
    \centering
    \includegraphics[width=1\linewidth]{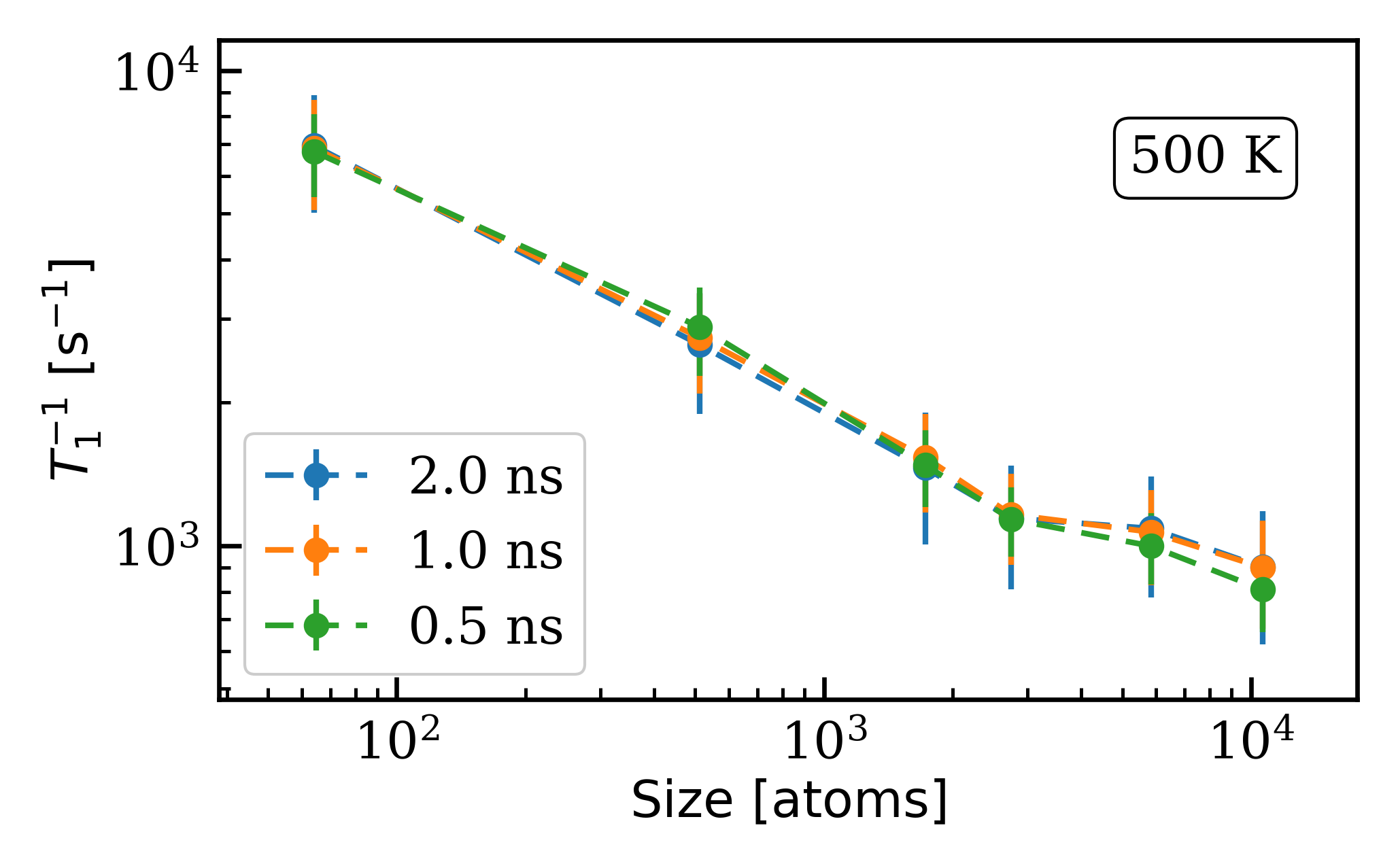}
    \caption{Convergence analysis of $T_1$ as a function of the number of atoms in the simulation cell and segment size used in the block analysis. All calculations are performed using $\mathrm{16~ns}$ NVE molecular dynamics  simulations with machine learned potentials at $\mathrm{500~K}$.}
    \label{fig: convergence test}
\end{figure}
%

To analyze  the effect of cell size and simulation times on our results, we focused on trajectories at $\mathrm{500~K}$ where acoustic vibrational modes are the most populated and large supercells are required to describe low frequency modes.
Error bars are estimated via block--analysis.
We considered supercells up to $10648$ atoms and divided the trajectories in segments of $\mathrm{0.5~ns}$, $\mathrm{1~ns}$ and $\mathrm{2~ns}$.
The complete convergence analysis presented in Fig.~\ref{fig: convergence test} indicates that at $500~\mathrm{K}$ there are significant finite size effects for diamond cells with less than $5000$ atoms and that the minimal block length is $1~\mathrm{ns}$. Therefore, thereinafter all numerical results are computed using simulation cells of at least $10648$ atoms and $\mathrm{1~ns}$-long blocks.

It is informative to decompose  Eq.s~\eqref{eq:spin-lattice relaxation}--\eqref{eq:decoherence relaxation} in secular and non--secular contributions. 
The secular terms, associated to $\omega_\delta=0$, describe eigenvalue fluctuations. The non secular ones are responsible for transitions between sub--levels of the triplet state of the NV center. 
Furthermore, Eqs.~\eqref{eq:spin-lattice relaxation}--\eqref{eq:decoherence relaxation} can be expanded in the frequency domain, $\omega$, via Fourier transforms: $1/{T_{1,2}}=\sum_{\omega_\delta}\lim_{\omega\to\omega_\delta}{T_{1,2}}^{-1}_{(\delta)}(\omega)$.
Our analysis highlights that spin--lattice relaxation depends only on non--secular contributions whereas $T_2$ is influenced by both~\cite{onizhuk_colloquium_2025}. 
The spin--lattice relaxation obtained in our formulation, being a pure non--secular process, is consistent with that derived within the Redfield formalism where $T_1$ is interpreted as the characteristic time associated to the decay of the diagonal elements of the density matrix of a two level system.
The secular part of $T_2$ is usually referred to as a {\it pure dephasing}  time, $T^{-1}_\Phi=\sum_{\omega_\delta=0}\lim_{\omega\to\omega_\delta}{T_{2}}^{-1}_{(\delta)}(\omega)$. The non--secular term arises from population exchanges, ${T^\prime_1}^{-1}=\sum_{\omega_\delta\neq0}\lim_{\omega\to\omega_\delta}{T_{2}}^{-1}_{(\delta)}(\omega)$, which contribute to decoherence, $1/T_2=T^{-1}_\Phi+{T^\prime_1}^{-1}$. 
\begin{figure}[t]
    \centering
    \includegraphics[width=1\linewidth]{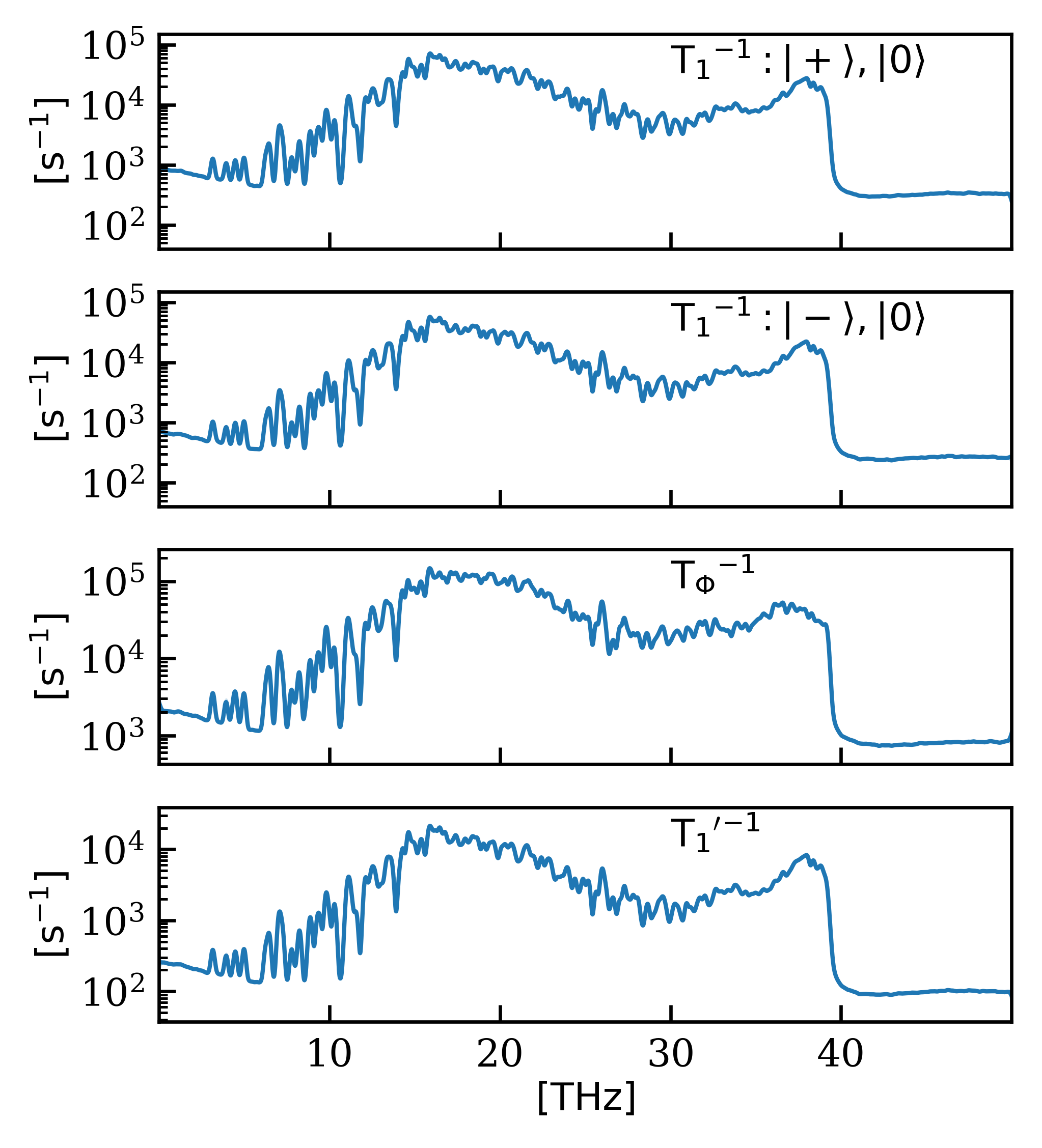}
    \caption{Frequency and transition decomposition of Eq.~\eqref{eq:spin-lattice relaxation} between $\ket{+}$--$\ket{0}$ and $\ket{-}$--$\ket{0}$ (upper and upper middle panels), of the pure dephasing contribution to $T_2$, $T_\Phi$ (lower middle panel) and of $T_1^\prime$ (lower middle panel). All calculations are performed using a $\mathrm{16~ns}$ NVE molecular dynamics  simulation with machine learned potentials of one NV center in a diamond cell of $10648$ atoms at $500~\mathrm{K}$.}
    \label{fig: contributions T1, T2}
\end{figure}
In Fig.~\ref{fig: contributions T1, T2}, we show the sub--levels and frequency decomposition of spin relaxation times computed at $\mathrm{500~K}$ for one NV center in diamond.
As expected, the contributions of the transitions $\ket{0}$--$\ket{\pm}$ to spin--lattice relaxation are of a similar magnitude, whereas pure dephasing , $T_\Phi^{-1}$, is about one order of magnitude larger than the population term, ${T^\prime_1}^{-1}$, in $T_2^{-1}$. 
Interestingly, the Fourier decompositions of Eq.s~\eqref{eq:spin-lattice relaxation}--\eqref{eq:decoherence relaxation} computed from ML MD simulations reveal the influence of low frequency lattice vibrations on the dynamics of the ZFS tensor. Indeed, in Fig.\ref{fig: contributions T1, T2} we do not observe vibrational peaks after the optical resonance of crystalline diamond at $\mathrm{40~THz}$.

\begin{figure}[t] 
    \centering
    \includegraphics[width=1\linewidth]{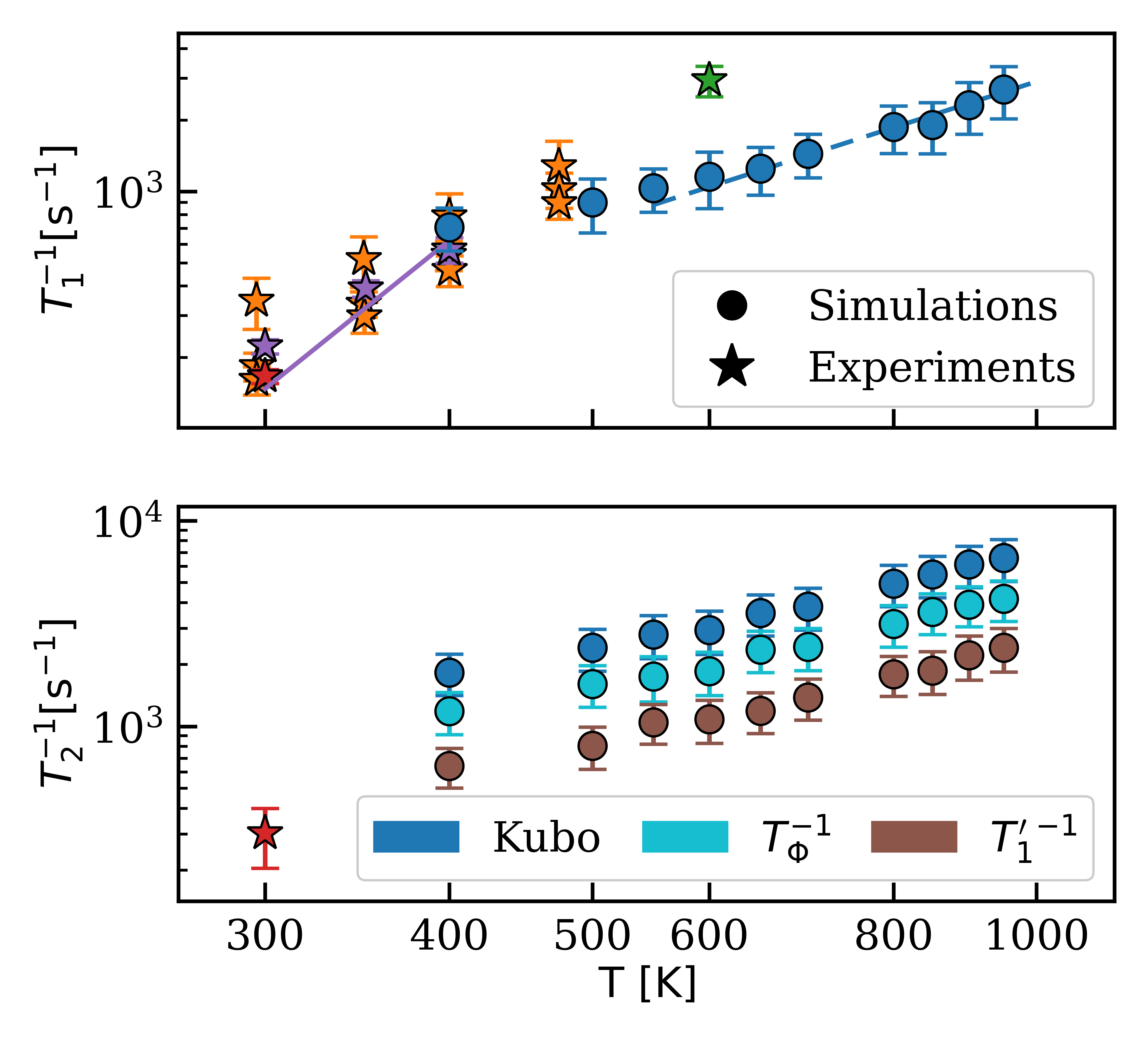}
    \caption{Spin-lattice relaxation, $T_1$, (upper panel) and decoherence time, $T_2$, (lower panel) of the NV center in diamond as a function of temperature. $T_1$ and $T_2$ are  evaluated using Eqs.~\eqref{eq:spin-lattice relaxation}--\eqref{eq:decoherence relaxation} from $\mathrm{16~ns}$ NVE molecular dynamics simulations of one NV center in a diamond cell of $10648$ atoms. We compare the numerical results of $T_1$ (blue circles) with our experimental results (purple stars),  and with those of Ref.~\cite{bar-gill_solid-state_2013} (red star), Ref.~\cite{jarmola_temperature-_2012} (orange stars) and Ref.~\cite{toyli_measurement_2012} (green star). We omit the result at  $\mathrm{1000~K}$, presented in Ref.~\cite{liu_coherent_2019}, as measurements were conducted on nanodiamonds, where relaxation mechanisms may substantially differ from those in bulk diamond.
    The purple solid line is the low temperature fit of $T_1^{-1}\approx a T^5$ for $T\leq \mathrm{400~K}$; the dotted blue line is the high temperature fit of $T_1^{-1}\approx bT^2$ for  $T\geq \mathrm{550~K}$.  We compare computed $T_2$ results (blue circles) with the measurements of Ref.~\cite{bar-gill_solid-state_2013}, and show the $T_\phi$ contribution (cyan circles) and $T_1^\prime$ (brown circles).}
    \label{fig:T1-2 vs T}
\end{figure}

Finally, we discuss the dependence of spin relaxation times on temperature, as obtained from ML MD simulations of one NV center in diamond, and from experiments conducted on purified diamond samples~\cite{SM}. 
The results are displayed in Fig.~\ref{fig:T1-2 vs T}, where we also present additional  experimental data from the literature~\cite{bar-gill_solid-state_2013,jarmola_temperature-_2012,toyli_measurement_2012}.  
As expected, we find that spin relaxation times present a significant temperature dependence, as they drastically decrease as temperature increases.
In our MD calculations, decoherence is caused by the interaction of the central spin with the lattice in the absence of magnetic noise and it is dominated by pure dephasing at all temperatures. The effect of temperature on $T_2$ at $\mathrm{500~K}$ is $T_2=\mathrm{414\pm87\mu s}$. We include the effect of the magnetic noise on $T_2$ using the Matthiessen's rule and computing the magnetic noise contribution with the Cluster Correlation expansion (CCE) method. Compared to the CCE results for the NV center in diamond at natural $\mathrm{[^{13}C]}$ abundance and $\mathrm{0~K}$~\cite{naguraprm2026}, we find that temperature effects at $\mathrm{500~K}$ reduce $T_2$ by a factor of 4.
The $T_1$ value computed from MD simulations at $\mathrm{600~K}$, $\mathrm{0.86\pm0.23~ms}$, is in fair agreement with measurements conducted at the same temperature, $\mathrm{340\pm50\mu s}$~\cite{toyli_measurement_2012}.
Instead, our computed $T_1$ at $\mathrm{500~K}$, $\mathrm{1.11\pm0.28~ms}$, is in quantitative agreement with previous experimental results, $\mathrm{0.94\pm0.11~ms}$ at $\mathrm{475~K}$~\cite{jarmola_temperature-_2012}. When considering 
 $100\%\mathrm{[^{13}C]}$ in diamond at $\mathrm{500~K}$, we found a moderate increase of $T_1$ to $\mathrm{1.26\pm0.27~ms}$, reflecting solely the effect of mass change in the sample.
The experimental measurements performed in this work spanning $\mathrm{300~K}$ to $\mathrm{400~K}$ allow us to unify the analysis of the spin lattice relaxation time as a function of temperature. Notably, at $\mathrm{400~K}$ our numerical result for $T_1$, $\mathrm{1.41\pm0.29~ms}$, is in excellent agreement with our experiments $\mathrm{1.75\pm0.22~ms}$, and previous measurements, $\mathrm{1.67\pm0.20~ms}$~\cite{jarmola_temperature-_2012}. We note that prior lattice dynamics calculations considering only two--phonons processes, while accurate at low temperatures, show only a fair agreement with experiments around $\mathrm{400~K}$~\cite{Mondal2023,cambria_temperature-dependent_2023}.

Under the assumption of Raman processes by spin-one-phonon interaction in second order, $T_1^{-1}$ at low temperatures follows the well known $T^5$ temperature behavior and at high temperatures a $T^2$ scaling law~\cite{Shrivastava1983}. In the case of diamond, whose Debye temperature is $\approx\mathrm{2000~K}$, the crossover temperature from the $T^5$ to $T^2$ scaling is found approximately at $\approx\mathrm{500~K}$. In Fig.~\ref{fig:T1-2 vs T} we show that  the experimental measurements at $T\leq\mathrm{400~K}$ are accurately fitted by $T_1^{-1}\approx aT^5$ (solid purple line) and those at $T\geq\mathrm{550K}$ by $T_1^{-1}\approx bT^2$ (dotted blue line). We note that the experimental result at $\mathrm{600~K}$ reported by  Ref.~\cite{toyli_measurement_2012} does not follow the expected $T^2$ scaling, when considered together with the other data. A possible explanation for the disagreement is the poor contrast and performance drop at these high temperatures.

{\it Discussion.}---In summary, we derived spin--lattice and decoherence relaxation times within Kubo linear response theory, describing both processes as the regression of two conserved quantities towards equilibrium, spin energy and transverse magnetization.
By interpreting these phenomena as responses to external driving fields, the use of Kubo LRT unifies the description of spin-lattice relaxation and decoherence in a single theoretical framework, as shown 
by Eqs.~\eqref{eq:spin-lattice relaxation}--\eqref{eq:decoherence relaxation}, where we expressed $T_1$ and the lattice contribution to $T_2$ in terms of equilibrium time-correlation functions of spin--lattice couplings.

We emphasize that the study of spin relaxation had so far  been limited to low temperatures, using the Redfield ME and CCE methods~\cite{redfield_theory_1965,lunghi_toward_2022,mariano_role_2025,onizhuk_colloquium_2025,onizhuk_bath-limited_2023,onizhuk_probing_2021,onizhuk_understanding_2024,naguraprm2026,Mondal2023,Xu2020,Xu2024,cambria_temperature-dependent_2023,Lunghi2017,Lunghi2023}. However, CCE techniques only account for the decoherence induced by the nuclear spins bath in the frozen lattice approximation~\cite{onizhuk_colloquium_2025,onizhuk_bath-limited_2023,onizhuk_probing_2021,onizhuk_understanding_2024,naguraprm2026}. Further, the Redfield ME relies on specific assumptions about one and two phonons processes\cite{Mondal2023,lunghi_toward_2022,mariano_role_2025,redfield_theory_1965,Lunghi2017,Lunghi2023,Xu2020,Xu2024,cambria_temperature-dependent_2023}, and
corrections of the Redfield ME at higher temperatures, where the harmonic approximation is no longer valid, are computationally intensive~\cite{chanda2026extendingspinlatticerelaxationtheory, Lunghi2017}.
Here, for the first time, we studied the coherence properties of the NV center in diamond at high temperatures combining Kubo LRT and MD simulations. Crucially, using ML methods to accelerate both MD simulations and the calculations of ZFS tensors,  we could perform ns--long simulations for cells with $\approx10000$ atoms. 
Our framework takes into account all phonon processes and anharmonicity.
Notably, in the phonon quasi--particle limit of a two level system, Eq.~\eqref{eq:spin-lattice relaxation} is equivalent to the second order Redfield ME. 



Our results show, as expected, that  $T_1$ depends only on population exchange within the sub--levels of the triplet state,  while $T_2$ is dominated by secular pure dephasing.
Our numerical predictions at $\mathrm{500~K}$ and $\mathrm{400~K}$ are in excellent agreement with our own  experiments and previous measurements~\cite{jarmola_temperature-_2012}.

Importantly, the framework presented here, based on Kubo LRT, is not restricted to ZFS Hamiltonians and can be extended to any system, spin-lattice interaction, and to any distribution of isotopic disorder.
Further it can be readily applied to any triplet color centers in the solid state or to molecular qubits.
In addition, thanks to the comprehensive use of NN models, our  MD approach can approximate any {\it ab--initio} level of theory via active learning techniques.  



\begin{acknowledgments}
{\it Acknowledgments.}---We gratefully acknowledge fruitful discussions with David M. Toyli, Stefano Villani, Swarnabha Chattaraj, Lien T. Le, Jonah Nagura, Alfredo Fiorentino and Stefano Baroni. The theoretical, experimental and computational work was supported by the Midwest Integrated Center for Computational Materials (MICCoM) as part of the Computational Materials Sciences Program funded by the U.S. Department of Energy (E. D., F.J.H and G. G.). This research used resources of the National Energy Research Scientific Computing Center (NERSC), a DOE Office of Science User Facility supported by the Office of Science of the U.S. Department of Energy under Contract No. ERCAP0036175 and resources of the University of Chicago Research Computing Center (RCC). Additional support for experimental validation was provided by Q-NEXT (M.M.M.); the Quantum Fellowship from the University of Chicago and the AFOSR MURI under award No. FA9550-23-1-0330 (Y.D.); and the U.S. Department of Energy, Office of Science, Basic Energy Sciences, Materials Sciences and Engineering Division through Argonne National Laboratory (B.P. and D.D.A.) under Contract No. DE-AC02-06CH11357.
\end{acknowledgments}

{\it Data availability}---The code, inputs and data that support the findings of this article are openly available at [DOIs].

\bibliography{main.bib}

\onecolumngrid
\section*{End Matter}
\twocolumngrid
\appendix
\setcounter{equation}{0}
\renewcommand{\theequation}{A\arabic{equation}}
{\it Appendix A: Derivation of Eq.~\eqref{eq:decoherence relaxation} and Eq.~\eqref{eq:spin-lattice relaxation}}---We consider a system described by an Hamiltonian $\mathcal{H}$,
\begin{align}
    \hh=\hh_0+\hh^\prime,
\end{align}
where $\hh_0$ is the reference Hamiltonian and $\hh^\prime$ is an  interaction term, $\comm{\hh_0}{\hh^\prime}\neq0$.
We study the evolution of an observable $\ha$, such that $\comm{\ha}{\hh_0}=0$, under an external perturbation that couples with $\ha$, $\hh_{ext}=\ha\lambda$.
Within linear order in the external perturbation, and in the limit of high temperatures, we can write~\cite{kubo_general_1954,kubo_statistical-mechanical_1957}:
\begin{align}\label{eq: kubo}
    \dot{A}\equiv\expval{\dot{\ha}}&
    =\beta\int_0^\infty\left\langle \dot{\ha}(t)\dot{\ha} \right\rangle\dd{t} \lambda
\end{align}
where $\beta = {1}\over{KT}$. In the interaction picture with respect to $\hh_0$ the correlation function in Eq.~\eqref{eq: kubo} is:
\begin{align}\label{eq: CAA}
    \left\langle\dot{\ha}(t)\dot{\ha}\right\rangle=&
    \left\langle U^\dagger_I(t)\left(\frac{i}{\hbar}\comm{\hh^\prime(t)}{\ha_0(t)}\right)U_I(t)\left(\frac{i}{\hbar}\comm{\mathcal{H}^\prime}{\ha}\right)\right\rangle
\end{align}
where on the left-hand side of Eq.~\eqref{eq: CAA} the operators are in the Heisenberg picture, while on the right-hand side of Eq.~\eqref{eq: CAA} operators are represented in the interaction picture: ${\ha_0}$ denotes the operator $\ha$ in the interaction picture and $U_I(t)$ is the interaction-picture time-evolution operator,
\begin{align}
    U_I(t)=1-\frac{i}{\hbar}\int_0^t\mathcal{H}^\prime(t^\prime)U_I(t^\prime)\dd{t^\prime}.
\end{align}
Within second order in the interaction $\hh^\prime$, we write:
\begin{align}
    \left\langle\dot{\ha}(t)\dot{\ha}\right\rangle=\left(\frac{i}{\hbar}\right)^2\left\langle 
    \comm{\mathcal{H}^\prime(t)}{\ha_0(t)}\comm{\mathcal{H}^\prime}{\ha}\right\rangle + \mathcal{O}({\mathcal{H}^\prime}^3).
\end{align}
Therefore, in the limit of high temperatures, at first order in the external perturbation and at second order in the interaction, Kubo LRT yields~\cite{kubo_general_1954}
\begin{align}
    \dot{A}&=\frac{\beta}{\hbar^2}\int_0^\infty\left\langle 
    \comm{{\ha_0}(t)}{\mathcal{H}^\prime(t)}\comm{\mathcal{H}^\prime}{\ha}\right\rangle\dd{t} \lambda.
\end{align}
which reduces to Eq.3 and Eq.4 in the main text when we substitute $\ha\equiv \hh_\mathrm{S}^0$ and $\ha\equiv\bm{\mathcal{M}}^\perp$, respectively.

\begin{figure*}[t]
    \centering
    \includegraphics[width=\textwidth]{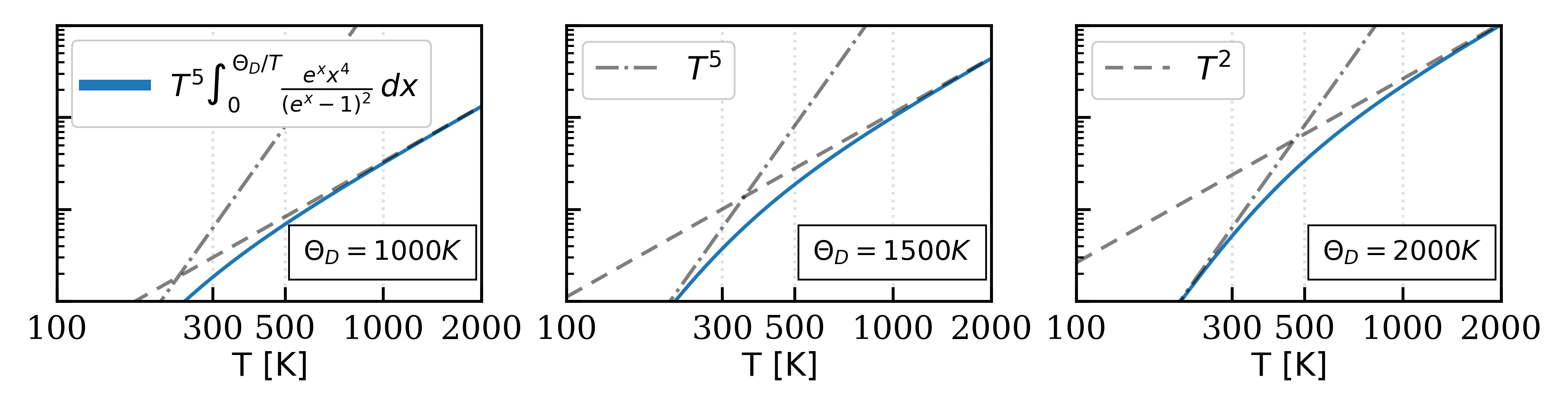}
    \caption{Eq.~\eqref{eq: shrivastava} as a function of temperature for  $\Theta_D=\mathrm{1000~K}, \mathrm{1500~K}, \mathrm{2000~K}$. The solid blue line represents the analytical result, the gray dashed and dash--dotted lines show the $T^2$ and $T^5$ scaling laws, respectively.}
    \label{fig:rahman}
    \centering
    \includegraphics[width=\textwidth]{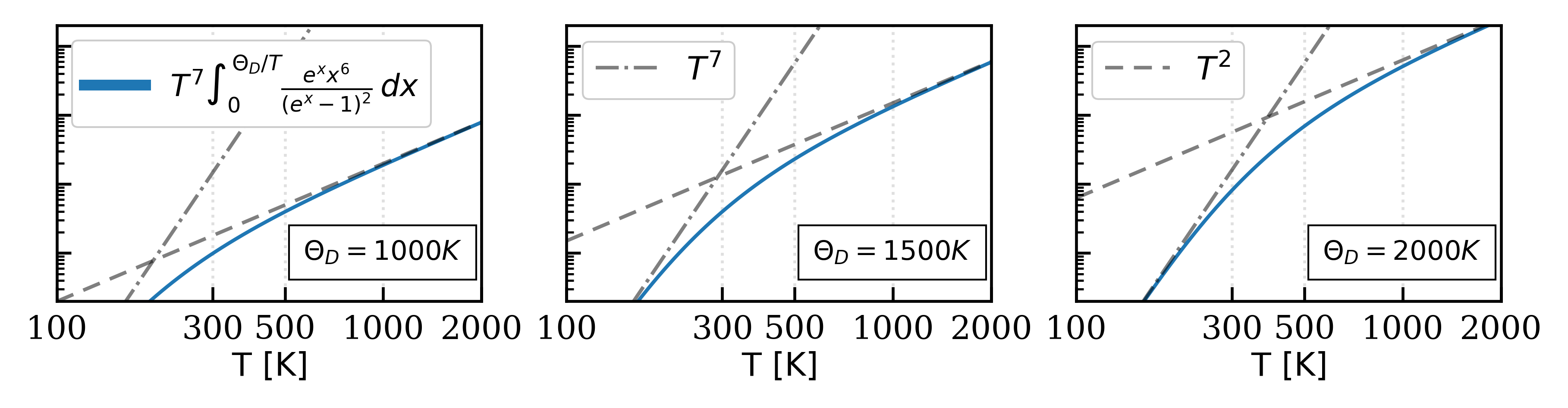}
    \caption{Eq.~\eqref{eq: shrivastava raman 2ph 1st} as a function of temperature for  $\Theta_D=\mathrm{1000~K}, \mathrm{1500~K}, \mathrm{2000~K}$. The solid blue line represents the analytical result, the gray dashed and dash--dotted lines show the $T^2$ and $T^7$ scaling laws, respectively.}
    \label{fig:rahman-2}
\end{figure*}

{\it Appendix B: Temperature dependence of Raman scattering by spin--one--phonon interaction and spin--two--phonon interaction }---Two phonon processes are the dominant decoherence mechanisms in the NV center in diamond~\cite{bar-gill_solid-state_2013,jarmola_temperature-_2012}. Specifically, for the NV center in diamond, the experimental $T^{-5}$ behavior of $T_1$ at low temperatures has been attributed to Raman processes by spin--one--phonon interaction at the second order~\cite{bar-gill_solid-state_2013,jarmola_temperature-_2012,Shrivastava1983,liu_coherent_2019}. In order to study Raman processes by spin--one--phonon interaction, we expand the spin--lattice interaction at first order in the displacement $Q_\mu$ as 
\begin{align}
    \hh_\mathrm{SL}^1\equiv\sum_\mu \left.\frac{\partial\hh_\mathrm{S}}{\partial Q_\mu}\right\vert_{0} \left(\mathcal{Q}_\mu + \mathcal{Q}_\mu^\dagger\right)
\end{align}
where $\mathcal{Q}_\mu$ is the creation operator of a phonon with energy $\hbar\omega_\mu$.
The probability for a spin state $a$ to be excited in a virtual state $c$ by absorption of a phonon, and subsequently for the level $c$ to decay to the desired state $b$ by emission of a phonon, $P_{ab}$, is~\cite{Shrivastava1983}
\begin{widetext}
\begin{align}\label{eq: Pab}
    P_{ab}=\frac{2\pi}{\hbar}\left\vert\frac{\bra{b, n_{\bm{k}}-1, n_{\bm{k}^\prime}+1}\hh_\mathrm{SL}^1\ket{c,n_{\bm{k}}-1, n_{\bm{k}^\prime}}\bra{c,n_{\bm{k}}-1, n_{\bm{k}^\prime}}\hh_\mathrm{SL}^1\ket{a, n_{\bm{k}}, n_{\bm{k}^\prime}}}{E_a-E_c-\omega_{\bm{k}}}\right\vert^2\times\delta(E_a-E_b-\omega_{\bm{k}}+\omega_{\bm{k}^\prime}).
\end{align}
\end{widetext}
In the Debye approximation and in the limit of $E_c-E_a\ll\omega_{\bm{k}}$ and $E_b-E_a\ll\omega_{\bm{k}}$, Eq.~\eqref{eq: Pab} reduces to~\cite{Shrivastava1983}
\begin{align}\label{eq: shrivastava}
    P_{ab}\propto \left(\frac{T}{\Theta_D}\right)^5\int_0^{\Theta_D/T} \frac{e^x x^4}{\left(e^x-1\right)^2}\dd{x},
\end{align}
where $\Theta_D$ is the Debye temperature of the material.
In Fig.~\ref{fig:rahman} we plot the right-hand side of Eq.~\eqref{eq: shrivastava}
as a function of temperature and show  that for low temperatures we recover the expected $T^5$ scaling law~\cite{Shrivastava1983}. At high temperatures, Eq.~\eqref{eq: shrivastava} shows $T_1^{-1}\propto T^2$~\cite{Shrivastava1983}. 
In diamond, $\Theta_D\approx\mathrm{2000~K}$, and the crossover temperature of Eq.~\eqref{eq: shrivastava} from $T^5$ to $T^2$ scaling is found at $\approx\mathrm{500~K}$.

Ref.~\cite{Mondal2023} and Ref.~\cite{cambria_temperature-dependent_2023} have investigated the role of Raman spin--two--phonon scattering on $T_1$. The Spin Hamiltonian is expanded at second order in the displacements:
\begin{align}
    \hh_\mathrm{SL}^2\equiv\sum_{\mu,\nu} \left.\frac{\partial^2\hh_\mathrm{S}}{\partial Q_\mu\partial Q_\nu}\right\vert_{0} \left(\mathcal{Q}_\mu + \mathcal{Q}_\mu^\dagger\right)\left(\mathcal{Q}_\nu + \mathcal{Q}_\nu^\dagger\right)
\end{align}
In the Debye approximation and in the limit of $E_b-E_a\ll\omega_{\bm{k}}$, the probability of scattering from the state $a$ to $b$ via emission and absorption of a phonon is:
\begin{align}\label{eq: shrivastava raman 2ph 1st}
    P_{ab}\propto \left(\frac{T}{\Theta_D}\right)^7\int_0^{\Theta_D/T} \frac{e^x x^6}{\left(e^x-1\right)^2}\dd{x}.
\end{align}
Fig.~\ref{fig:rahman-2} displays the right-hand side of Eq.~\eqref{eq: shrivastava raman 2ph 1st}
as a function of temperature and at low temperatures we recover the expected $T^7$ scaling law~\cite{Shrivastava1983}. As for the Raman spin--one--phonon scattering in second order, at high temperatures, Eq.~\eqref{eq: shrivastava raman 2ph 1st} shows $T_1^{-1}\propto T^2$~\cite{Shrivastava1983}. 

\end{document}


\linenumbers

\title{Supplemental Material for ``Spin Dynamics from Atomistic Quantum Simulations''}


\author{Enrico Drigo\,\orcidlink{0000-0002-1797-2987}}
\email[]{endrigo@uchicago.edu}
\affiliation{Pritzker School of Molecular Engineering, University of Chicago, Chicago, Illinois 60637, USA}
\author{Marquis M. McMillan\,\orcidlink{0009-0009-4664-6556}}
\affiliation{Pritzker School of Molecular Engineering, University of Chicago, Chicago, Illinois 60637, USA}
\author{Benjamin Pingault\,\orcidlink{0000-0003-1259-2747}}
\affiliation{Q-NEXT, Argonne National Laboratory, Lemont, Illinois 60439, USA}
\affiliation{Materials Science Division, Argonne National Laboratory, Lemont, Illinois 60439, USA}
\affiliation{Pritzker School of Molecular Engineering, University of Chicago, Chicago, Illinois 60637, USA}
\author{Yinan Dong\,\orcidlink{0000-0002-5996-642X}}
\affiliation{Pritzker School of Molecular Engineering, University of Chicago, Chicago, Illinois 60637, USA}
\author{F. Joseph Heremans\,\orcidlink{0000-0003-3337-7958}}
\affiliation{Pritzker School of Molecular Engineering, University of Chicago, Chicago, Illinois 60637, USA}
\affiliation{Q-NEXT, Argonne National Laboratory, Lemont, Illinois 60439, USA}
\affiliation{Materials Science Division, Argonne National Laboratory, Lemont, Illinois 60439, USA}
\author{David D. Awschalom\,\orcidlink{0000-0002-8591-2687}}
\email{awsch@uchicago.edu}
\affiliation{Pritzker School of Molecular Engineering, University of Chicago, Chicago, Illinois 60637, USA}
\affiliation{Q-NEXT, Argonne National Laboratory, Lemont, Illinois 60439, USA}
\affiliation{Materials Science Division, Argonne National Laboratory, Lemont, Illinois 60439, USA}
\affiliation{Department of Physics, University of Chicago, Chicago, Illinois 60637, USA}
\author{Giulia Galli}
\email[]{gagalli@uchicago.edu}
\affiliation{Pritzker School of Molecular Engineering, University of Chicago, Chicago, Illinois 60637, USA}
\affiliation{Department of Chemistry, University of Chicago, Chicago, Illinois 60637, USA}
\affiliation{Materials Science Division, Argonne National Laboratory, Lemont, Illinois 60439, USA}


\date{\today}


\maketitle

\section{Computational details}
All first principle (FP) calculations reported in this work are conducted at the DFT/PBE spin-polarized level of theory, constrained total magnetization $2$, $\mathrm{100~Ry}$ wavefunction kinetic energy cutoff , $10^{-12}$ self-consistent convergence threshold and ONCV pseudopotentials, using the Quantum ESPRESSO code~\cite{doi:10.1021/acs.jctc.3c00249,SCHLIPF201536,PhysRevB.88.085117,PhysRevLett.77.3865}.
The FP and machine learning (ML) molecular dynamics (MD) simulations are carried out using the Velocity Verlet algorithm with a time step of $\mathrm{1~fs}$ as implemented in Quantum ESPRESSO and in \texttt{LAMMPS} compiled with the \texttt{Symmetrix} package, respectively~\cite{THOMPSON2022108171,doi:10.1021/jacs.4c07099,10.1063/5.0297006, doi:10.1021/acs.jctc.3c00249}. The NVT FP MD simulations are performed using the Quantum ESPRESSO code and the Nosé-Hoover thermostat~\cite{doi:10.1021/acs.jctc.3c00249}. The NVT and NPT ML MD simulations are run using the Nosé-Hoover thermostat and barostat, respectively, and  \texttt{LAMMPS}~\cite{THOMPSON2022108171}.


The MACE machine learning interatomic potentials (MLIP) datasets were initially generated from the FP trajectories, extracting   $\approx6500$  and $\approx4000$ configurations for training and test set, respectively~\cite{NEURIPS2022_4a36c3c5}.
The MLIP training dataset was further refined using an active learning (AL) procedure. 
During the AL process, we estimate the uncertainty on the MD configurations as the maximum deviation of the predicted atomic forces among an ensemble of 4 independently trained MLIPs.
If the uncertainty of a configuration lays between $\mathrm{50~meV/\text{\AA}}$ and $\mathrm{100~meV/\text{\AA}}$, the structure is considered a candidate and added to the training dataset~\cite{bidoggia2026aiidatrainspotautomatedtrainingneuralnetwork}.
The AL loop converges when in the MD trajectories there are no more candidates. 
Specifically, we perform $\mathrm{100~ps}$-long NVT and NPT ML MD simulations to explore the phase diagram of the NV center in diamond in the same thermodynamic conditions as those sampled by the FP MD trajectories.
Finally, the training set contains $\approx8000$ samples. 
The MACE model architecture is $\texttt{64x0e}$ with a radial cutoff of $3.5~\text{\AA}$.  
The forces' root mean squared error (RMSE) computed on the test set is $\mathrm{24~meV/\text{\AA}}$.

The ZFS NN architecture consists of two messages, $50$ channels and a radial cutoff of $3.5$\AA, the initial training and test datasets are generated sampling $\approx1000$ and $\approx500$ structures from the FP MD trajectories, respectively.
The active learning loop for the ZFS NN is the same as for the MACE MLIP, with the notable distinction that the uncertainty is estimated from the standard deviation of the predicted ZFS tensors over an ensemble of independently trained ZFS NN models.
The final ZFS NN training dataset contains  $\approx3000$ structures. 
The RMSE per spherical component over the test set is $\mathrm{60~MHz}$.

\section{Sample Details}

The sample used for experimental comparison was grown with a 3 min exposure to a 1 sccm $^{15}$N$_2$ flow, corresponding to an estimated depth of $\sim 50$ nm. Nitrogen $\delta$-doping was introduced via isotopically enriched $^{15}$N$_2$ gas (99.99\% chemical purity, 99.9 at.\% isotopic enrichment) during growth. Comparison to a subsequent sample grown under similar conditions with a 10 sccm flow \cite{PhysRevMaterials.8.026204} indicates an approximately fivefold reduction in nitrogen incorporation, corresponding to an estimated concentration of [N] $\sim 0.08$--$0.8$ ppm. This low incorporation level is consistent with minimal additional magnetic noise.

\section{Experimental Details}

Our custom-built confocal microscope is designed to image and probe NV centers in diamond. We use a linearly unpolarized 532~nm laser (H\"{U}BNER Photonics Cobolt 06-MLD), pulsed via an external acousto-optical modulator (ISOMET 1205C) for time-domain measurements. The laser is focused onto the diamond sample using a top-down 100x/0.9 NA objective (NIKON LU Plan Fluor EPI A Pol) with a numerical aperture (NA) of 0.9. NV photoluminescence is separated from the excitation path using a 550~nm longpass dichroic mirror and spectrally filtered through a 655~nm longpass filter before detection with an avalanche photodiode (PerkinElmer SPCM-AQR-16). Spatial scanning is performed using a fast-steering mirror (Newport FSM-300) to locate and address single NV centers across the sample. Isotopically purified NV centers are confirmed via optically detected magnetic resonance (ODMR) through observation of the $^{15}$N hyperfine splitting. Coherent spin rotations are driven by a vector signal generator (Stanford Research Systems SG396), with microwave signals amplified (Mini-Circuits ZHL-16W-43+) and delivered to the sample through a 10~$\mu$m diameter gold wire bonded across the diamond surface. A permanent magnet (K\&J Magnetics) is mounted on a custom goniometer with a two-axis translation stage (Newport MFA-CC) for precise field alignment along the NV axis. All instrument timing is synchronized using an arbitrary waveform generator (Swabian Instruments Pulse Streamer 8/2). Temperature control between 300~K and 400~K is managed using a custom-built PID heater platform. The resistive heat source (TEMPCO Strip Heater, 60 W) is rated stable up to 430~K. A thermocouple is mounted directly to the sample stage as close as possible to the diamond surface. Given diamond's exceptionally high thermal conductivity of approximately 1000-2200~W~m$^{-1}$~K$^{-1}$ \cite{Kidalov2009Diamond}, we treat the sample as thermally homogeneous and estimate a temperature accuracy of better than 1~K across the probed region. Temperature setpoints are regulated using a PID controller (BriskHeat PID Temperature Controller).

To probe the spin-lattice relaxation time $T_1$ of the NV center at zero applied field, we performed all-optical relaxation measurements. The pulse sequence consists of an initialization pulse, a variable free-evolution interval $\tau$, and a readout pulse, with no microwave manipulation. The NV spin is first polarized into the $m_s = 0$ ground state via a green laser pulse of sufficient duration ($\sim$3--5~$\mu$s). After a variable dark interval $\tau$ during which the spin evolves freely under the influence of the thermal bath, a second green pulse both reads out the spin-state-dependent photoluminescence and re-initializes the system. Because the NV photoluminescence intensity is higher for $m_s = 0$ than for $m_s = \pm1$, the contrast decays as the spin population thermalizes toward a mixed state over the timescale $T_1$. Fitting the normalized contrast as a function of $\tau$ to a single-exponential decay yields $T_1$ directly, without requiring any microwave control.

\bibliography{SI.bib}